\documentclass[preprint,aps]{revtex4}
\usepackage{amsfonts}
\usepackage{graphicx}
\usepackage{bm}

\begin{document}

\title{Detecting Noncommutative Phase Space by Aharonov-Bohm Effect}
\author{Shi-Dong Liang}
\altaffiliation{Corresponding author: Email: stslsd@mail.sysu.edu.cn}
\author{Haoqi Li}
\author{Guang-Yao Huang}
\affiliation{State Key Laboratory of Optoelectronic Material and Technology, and
Guangdong Province Key Laboratory of Display Material and Technology,
School of Physics and Engineering, Sun Yat-Sen University, Guangzhou, 510275,
People's Republic of China}
\date{\today }

\begin{abstract}
Noncommutative phase space plays an essential role in
particle physics and quantum gravity in the Planck's scale. However, there has not been a direct
experimental evidence or observation to demonstrate the existence of
noncommutative phase space. We study quantum ring in noncommutative phase space based on
the Seiberg-Witten map and give the effective magnetic potential and field coming from the noncommutative phase space, which induces the persistent current in the ring. We introduce two variables as two signatures to detect the noncommutative phase space and propose an experimental scheme to detect the noncommutative phase space
as long as we measure the persistent current and the external magnetic flux.
\end{abstract}

\maketitle



\section{Introduction}
The noncommutative concepts in physics can trace back to the momentum
algebra. These concepts are upgrated to any pair of conjugate variables
called as Heisenberg's algebra in quantum mechanics. It implies that the
phase space of elementary particles has an intrinsic uncertainty and
particles has a non-local behavior in microscopic world. This idea is
generalized to noncommutative space $\left[ \widehat{x}_{\mu },\widehat{x}%
_{\nu }\right] =i\theta _{\mu \nu }$ to improve the renormalizability
properties of an elementary-particle theory at short distances and to
understand the singularity in elementary-particle physics. \cite{Douglas}.
Especially, the non-locality of gravity leads to a lot of theories based on
noncommutative space, such as quantum gravity, string and noncommutative
gauge theories.\cite{Douglas}

In fact, the noncommutative phase-space behavior also occurs in condensed
matter physics.\cite{Gamboa,Ho} It has been found that there is an analogy between the
Landau's levels of the two-dimensional electron gas in the presence of
magnetic field and the free particle system in noncommutative phase space.%
\cite{Delduc} Quantum Hall effect and its topological invariance can be also
understood by the noncommutative phase-space language.\cite{Basu,Gamboa}
It implies that the noncommutative-space phenomenon can occur beyond the Planck's scale even the
physics in noncommutative space has not been understood completely.

Interestingly, Connes and Rieffel construct a mathematical formulation on
noncommutative geometry that is relevant to noncommutative space physics
and as a mathematical language of noncommutative space physics.\cite{Connes}

However, even though many phenomena in particle physics, gravity and
condensed matter physics are relevant to or described by the noncommutative
space or phase-space language, there has not been direct experimental
evidences to demonstrate the existence of noncommutative geometry or space.
The difficulty to find direct experimental or observation evidence of the
existence of noncommutative space is because the noncommutative space
phenomena are predicted occurring in the Planck's scale in particle physics
and gravity.

Recently, Chaichian et. al. discussed the Aharonov-Bohm (AB) and
Aharonov-Casher (AC) effects in noncommutative space and gave an AB and AC
phase shifts coming from the noncommutative space $\theta $.\cite{Chaichian, Li}
But they did not give an experimental scheme to detect the noncommutative effect.

In this paper, we apply the Seiberg-Witten map to study the AB effect in
noncommutative phase space. The noncommutative momenta lead to an
effective magnetic flux that induces the persistent current in a mesoscopic ring.
We propose a scheme to detect the effective magnetic flux coming from the noncommutative phase
space even the effect is very weak. Since the persistent
current and magnetic flux in mesoscopic rings can be implemented by
nanotechnology, it can be expected to detect the noncommutative phase space
effect by this scheme.

\section{Noncommutative phase spaces}
The positions and momenta in quantum mechanics follow a noncommutative
relationship,

\begin{eqnarray}
\left[ x_{i},p_{j}\right] &=&i\hbar \delta _{ij}\text{ \ } \\
\left[ x_{i},x_{j}\right] &=&\left[ p_{i},p_{j}\right] =0  \label{xp1}
\end{eqnarray}%
where $i,j=1,2,3$. This is called as a canonical commutative relation or
Heisenberg-Weyl algebra. This noncommutative algebra can be generalized to \cite{Seiberg}

\begin{eqnarray}
\left[ \widehat{x}_{i},\widehat{p}_{j}\right] &=&i\hbar \text{ } \\
\left[ \widehat{x}_{i},\widehat{x}_{j}\right] &=&i\theta _{ij} \\
\left[ \widehat{p}_{i},\widehat{p}_{j}\right] &=&i\widetilde{\theta }_{ij}%
\label{xp2}
\end{eqnarray}%
where $\theta _{ij}$ and $\widetilde{\theta }_{ij}$ are antisymmetric real
constant $3\times 3$ matrices. These noncommutative algebra are equivalent
to the Weyl-Moyal correspondence in field theories on noncommutative spaces
\cite{Douglas} and can be realized by the Seiberg-Witten map, \cite{Seiberg}
\begin{eqnarray}
\widehat{x}_{\mu } &=&a_{\mu \nu }x_{\nu }+b_{\mu \nu }p_{\nu } \\
\widehat{p}_{\mu } &=&c_{\mu \nu }x_{\nu }+d_{\mu \nu }p_{\nu }
\label{xp3}
\end{eqnarray}
between the canonical commutative relation and noncommutative phase space.
For two-dimensional system, the Seiberg-Witten map can be reduced to
\begin{equation}
\left(
\begin{array}{c}
\hat{x} \\
\widehat{y} \\
\hat{p}_{x} \\
\hat{p}_{y}%
\end{array}%
\right) =\left(
\begin{array}{cccc}
\alpha & 0 & 0 & -\frac{\theta }{2\alpha \hbar } \\
0 & \alpha & \frac{\theta }{2\alpha \hbar } & 0 \\
0 & \frac{\widetilde{\theta }}{2\alpha \hbar } & \alpha & 0 \\
-\frac{\widetilde{\theta }}{2\alpha \hbar } & 0 & 0 & \alpha%
\end{array}%
\right) \left(
\begin{array}{c}
x \\
y \\
p_{x} \\
p_{y}%
\end{array}%
\right)
\end{equation}
with
\begin{equation}
\theta \widetilde{\theta }=2\hbar ^{2}\alpha ^{2}\left( 1-\alpha ^{2}\right)
\end{equation}

In general, when $\alpha \neq 1$ means $\theta \widetilde{\theta }\neq 0$.
Namely $\theta \neq 0$ and $\widetilde{\theta }\neq 0$, which is a general
noncommutative phase space. When $\alpha =1,$ it means $\theta \widetilde{%
\theta }=0$. It contains three cases: (1) $\widetilde{\theta }=0$ and $%
\theta \neq 0$. It is noncommutative space; (2) $\theta =0$, and $%
\widetilde{\theta }\neq 0$. It is a noncommutative momenta; (3) $\theta =0$
and $\widetilde{\theta }=0$. It reduces to the canonical commutative
relation in quantum mechanics.

For the 2D free electron in noncommutative phase space, the Hamiltonian can
be written as
\begin{equation}
H_{nc}=\frac{1}{2m}\left( \widehat{p}_{x}^{2}+\widehat{p}_{y}^{2}\right) =%
\frac{1}{2m^{\ast }}\left[ \left( p_{x}+eA_{x}\right) ^{2}+\left(
p_{y}+eA_{y}\right) ^{2}\right]
\end{equation}
where $m^{\ast }=m/\alpha $\ is the effective mass in noncommutative phase
space. The effective vector potential is
\begin{equation}
A_{x}=\frac{\widetilde{\theta }}{2e\alpha ^{2}\hbar }y;\text{ \ \ \ }A_{y}=-%
\frac{\widetilde{\theta }}{2e\alpha ^{2}\hbar }x
\end{equation}%
and the effective magnetic field

\begin{equation}
B_{z}=\frac{\widetilde{\theta }}{e\alpha ^{2}\hbar }
\end{equation}
It can be seen that the free electron in the 2D noncommutative phase space is equivalent to electron in a effective magnetic field induced by the noncommutative phase space effect. It implies that we can demonstrate the existence of noncommutative phase space once we can measure the effect of the effective magnetic field or potential coming from the noncommutative phase space.

\section{Quantum ring and persistent current}

We consider a one-dimensional ring in an external magnetic field along the
axis of the ring. The magnetic field is constant inside $r_{c}<R$ (ring
radius) such that the electrons are in the field-free region and the
electron states depend only on the total magnetic flux in the ring. In the
polar coordinate system, $x=R\cos \varphi ,\qquad y=R\sin \varphi$, the
Hamiltonian in noncommutative phase space is written as

\begin{equation}
H_{nc}=-\frac{\hbar ^{2}}{2m^{\ast }R^{2}}\left[ \frac{\partial }{\partial
\varphi }+i\left( \frac{\phi }{\phi _{0}}-\frac{\phi _{nc}}{\phi _{0}}%
\right) \right] ^{2}-\frac{3\hbar ^{2}}{8m^{\ast }R^{2}}\frac{\phi _{nc}^{2}%
}{\phi _{0}^{2}}
\end{equation}
where $\phi _{nc}=\frac{2\pi R^{2}\widetilde{\theta }}{e\hbar \alpha ^{2}}$
is an effective magnetic flux coming from the noncommutative phase space
and $\phi _{0}=\frac{h}{e}$ is the flux quanta. $\phi$ is the external
magnetic flux in the ring. For convenience, we introduce the dimensionless
magnetic flux, $f_{nc}\equiv \frac{\phi _{nc}}{\phi _{0}}$ and $f=\frac{\phi
}{\phi _{0}}$, the Hamiltonian of quantum ring can be rewritten as
\begin{equation}
H_{nc}=-\varepsilon _{0}\left[ \frac{\partial }{\partial \varphi }+i\left(
f-f_{nc}\right) \right] ^{2}-\frac{3\varepsilon _{0}}{4}f_{nc}^{2}
\end{equation}
where $\varepsilon _{0}\equiv \frac{\hbar ^{2}}{2m^{\ast }R^{2}}$. The
eigenenergy can be obtained by
\begin{equation}
E_{n}=\varepsilon _{0}\left( n+f-f_{nc}\right) ^{2}-\frac{3\varepsilon _{0}}{4}f_{nc}^{2}
\label{Eign}
\end{equation}
where $n=0,\pm 1,\pm 2,...$. The corresponding wave function is $\psi
_{n}(\varphi )=\exp \left( in\varphi \right) $. It can be seen that the
effective magnetic flux coming from the noncommutative space modifies the
eigenenergy levels. Since $n=0,\pm 1,\pm 2,..$, and
notice that the eigenenergies in Eq. (\ref{Eign}) are invariant under $f-f_{nc}\rightarrow f-f_{nc}+1$, we can consider only the domain of $f-f_{nc}$ within $\left[ -\frac{1}{2},\frac{1}{2}\right]$ (the first Brillouin flux zone).\cite{Daniel,Huang}
Suppose there are $N$ electrons in the ring, and they occupy the energy
levels in low temperature, using $\sum_{n=1}^{k}n=\frac{k(k+1)}{2}$, and $\sum_{n=1}^{k}n^2=\frac{k(k+1)(2k+1)}{6}$, notice that for $N=2k+1$, the ground-state energy is $E_{g}=\sum_{n=0,\pm1,\pm2,...}^{\pm k}E_{n}$, and for $N=2k$, $E_{g}=\left(\sum_{n=0,1,2,...}^{k-1}+\sum_{n=-1,-2,...}^{-k}\right)E_{n}$,
the ground-state energy of the ring is obtained

\begin{equation}
E_{g}=\varepsilon _{0}\left\{
\begin{array}{cc}
\frac{N^{3}-N}{12}+N\left[ (f-f_{nc})^{2}-\frac{3}{4}f_{nc}^{2}\right] &
for\  N=2k+1 \\
\frac{N^{3}+2N}{12}-N(f-f_{nc})+N\left[ (f-f_{nc})^{2}-\frac{3}{4}f_{nc}^{2}%
\right] & for\  N=2k%
\end{array}%
\right.
\label{Eg}
\end{equation}
Moreover, the ground-state energy is symmetric for $f-f_{nc}=0$, we can restrict our attention to a half of the first Brillouin flux zone, $\left[0,\frac{1}{2}\right]$.
The persistent current in the ground state is defined by $J=-\frac{\partial
E_{g}}{\partial \phi }$ and it can be obtained

\begin{eqnarray}
J=J_{0}\left\{
\begin{array}{cc}
-2N\frac{\phi }{\phi _{0}}\left( 1-\frac{\phi _{nc}}{\phi }\right) & for\  N=2k+1
\\
N-2N\frac{\phi }{\phi _{0}}\left( 1-\frac{\phi _{nc}}{\phi }\right) & for\  N=2k%
\end{array}%
\right.
\label{Jphi}
\end{eqnarray}
where $J_{0}=\frac{e}{h}\varepsilon_{0}$. The persistent current depends on the external magnetic flux and the effective magnetic flux coming from the noncommutative phase space. The relationship between the persistent current and the magnetic flux in Eq. (\ref{Jphi}) provides a way to reveal the noncommutative phase space effect coming from the noncommutative phase space.

\section{Signatures of noncommutative-phase-space effect}
In order to detect the noncommutative-phase-space effect experimentally, we introduce two variables
defined by
\begin{eqnarray}
\lambda&\equiv& \frac{\partial }{\partial \phi }\left(\frac{J}{\phi}\right) \\
\sigma&\equiv& \frac{\partial }{\partial \phi }\left(\frac{J-NJ_{0}}{\phi }\right)
\end{eqnarray}
as two signatures to detect the noncommutative-phase-space effect experimentally. Thus, we get

\begin{eqnarray}
\lambda=\left\{
\begin{array}{cc}
-2NJ_{0}\frac{f_{nc}}{\phi ^{2}}  & for\  N=2k+1 \\
-NJ_{0}\left(1+2f_{nc}\right)\frac{1}{\phi^{2}}  & for\  N=2k%
\end{array}
\right.
\end{eqnarray}
and
\begin{eqnarray}
\sigma=\left\{
\begin{array}{cc}
NJ_{0}\left(1-2f_{nc}\right)\frac{1}{\phi^{2}}  & for\  N=2k+1 \\
-2NJ_{0}\frac{f_{nc}}{\phi ^{2}}  & for\  N=2k%
\end{array}
\right.
\end{eqnarray}

It can be seen that when there exists the noncommutative phase space the
behaviors of $\lambda\sim\phi $ and $\sigma\sim
\phi $ are $\frac{\pm1}{\phi^{2}}$ divergence in the region of small external
magnetic fluxes.

For given parameter $\widetilde{\theta }\le 1.76\times
10^{-61}[kg]^{2}[m]^{2}[s]^{-2}$ \cite{Bastos} and a mesoscopic
ring with radius $R=1\mu m$, $f_{nc}\le\frac{2\pi R^{2}\widetilde{\theta }}{%
e\hbar \alpha ^{2}h/e}=\frac{R^{2}\widetilde{\theta }}{\hbar ^{2}\alpha ^{2}}%
=\allowbreak 1.5828\times 10^{-5}$.
Suppose that the effective electron number in the ring is about $10^4\sim10^{5}$,
we show some theoretical predictions in Figures 1 and 2 based on this idea and parameters.

The figure 1 shows the theoretical prediction of the $\lambda
\sim\phi$ behavior in nano-ring with radius $R=1\mu m$ with odd-electron number.
It can be seen that the $\lambda$ is divergent in the small $\phi$ range.
In the inset of Fig.1 (a), we plot $\lambda\sim\phi$ by the log-log scale, which is linear that can be used as
standard of $-1/\phi^{2}$ behavior. Similarly in Fig.1 (b), we
shows the behavior of $\sigma\sim\phi$ with its log-log scale in the inset of Fig.1 (b).
It can be seen that $\lambda$ and $\sigma$ for odd electron number rings have a similar behavior, but the sign is different.

For the even-electron number rings, the behaviors of $\lambda$ and $\sigma$ are similar and have only a two-order difference, which are shown in Fig.2 (a) and (b). The linear behaviors of $\lambda\sim\phi$  and the log-log scale of $\sigma\sim\phi$ are also used a standard of the test the $1/\phi^{2}$ divergent behavior.

\section{Scheme for detecting noncommutative-phase-space effect}

We can propose an experimental scheme to demonstrate explicitly
the noncommutative phase space based on above results. The basic steps are

(1) set up a mesoscopic ring system with an external magnetic field;

(2) measuring the persistent current $J$ versus the external magnetic flux $\phi $;

(3) by using the numerical interpolation and derivative techniques to
calculate $\lambda$ and $\sigma$ by estimating the electron number $N$;

(4) plot $\lambda$ versus $\phi$ and $\sigma$ versus $\phi$,
if we can obtain a qualitative the behaviors of $\lambda$ versus $\phi$ or $\sigma$ versus $\phi$ in Figs 1 and 2, it demonstrates the existence of the noncommutative phase space.

It should be remarked that the resolution of the numerical estimations in
Figs (1) and (2) are based on the parameters of the Planck's scale and nano
size of the ring. The comparison between the experimental data and the theoretical prediction in Figs 1 and 2 are meaningful and valid as long as in qualitative level because the theoretical prediction is based on the noncommutative space parameter, $\widetilde{\theta }$, and electron number $N$, but we do not know the exact values of these parameter.
In other words, the value $\widetilde{\theta }$ given in the numerical results is not crucial for the conclusion. The variance of $\widetilde{\theta }$ shifts just the curves in Figs. (1) and (2) up or down because $f_{nc}$ is proportional to $\widetilde{\theta }$.
Once we obtain $\pm1/\phi^{2}$ divergent curve we can demonstrates the existence of the noncommutative phase space. This scheme provides an efficient experimental way to detect the noncommutative space, especially estimating the noncommutative space parameter $\widetilde{\theta }$ when we can estimate the electron number. Moreover, we can also know the odd-even property of electron number from the basic behavior of
$\lambda$ and $\sigma$ in Figs 1 and 2.

The main conclusion can be summarized in the following criterion.\\
{\bf Criterion}: if one of the following two cases occurs it implies the existence of the noncommutative phase space:\\
(1) $\lambda\sim\phi$ is $-1/\phi^2$ divergent but $\sigma\sim\phi$ is $1/\phi^2$ divergent (see Fig. 1);\\
(2) both $\lambda\sim\phi$ and $\sigma\sim\phi$ are  $-1/\phi^2$ divergent and $\lambda<\sigma$
(see Fig. 2).

If (1) occurs it implies that there is odd number of electrons in the ring and if (2) occurs that infers the ring having even number of electrons.

In fact, the persistent current in mesoscopic ring have
been studied both theoretically and experimentally in past two decades.
\cite{Huang, Buttiker, Levy, Mailly} Buttiker first
predicted that the persistent current occurs in mesoscopic ring and
oscillates with an AB flux.\cite{Buttiker} It has been found that the amplitude of the
persistent current reaches $\left( 10^{-2}\sim 2\right) ev_{F}/2\pi
R $ where $v_{F}$ is the electronic velocity at Fermi level in the $Cu$
multi-ring system, an isolate $Au$ ring and $GaAs/Al_{x}Ga_{1-x}As$ in the
diffusive region at low temperature.\cite{Levy,Chandrasekhar, Mailly},
which agrees with the theoretical prediction. This
nanotechnology provides an efficient way to detect the noncommutative phase
space. We suggest to rerun the persistent-current experiment to measure the
persistent versus the external magnetic flux data to detect the noncommutative phase
space.

On the other hand, Carroll et. al. studied the noncommutative field theory and Lorentz violation. They give an upper bound of the noncommutative parameter, $\widetilde{\theta }\le\left(10TeV\right)^{-2}$.\cite{Carroll} Falomir et. al. proposed a scheme to explore the spatial noncommutativity of the scattering differential cross section by the AB effect.\cite{Falomir} It relies on the particle physics experiment involving energies between $200$ and $300$ GeV for $\widetilde{\theta}\le\left(10TeV\right)^{-2}$ and estimating the typical order of the cross section for neutrino events $\sim10^{-3}$. \cite{Falomir}
Obviously, this experimental scheme is much more difficult to be implemented than our scheme
because our scheme involves $eV$ energy scale and nanoscale physics.

In fact, the noncommutativity is not at all clear, it should be worth studying and exploring from different aspects and different energy scales even though the concept of noncommutativity originates from the Planck scale physics. Actually, many phenomena in condensed matter physics show the characteristics of noncommutative space in nonrelativistic quantum mechanics, such as an analogy between the Landau's levels of two-dimensional electron gas in the presence of magnetic field and the free electron in noncommutative phase space,\cite{Gamboa,Ho}, as well as quantum Hall effect.\cite{Basu}  So, it should be expected there exists possibility to detect the noncommutativity especially in the nanoscale condensed matter physics.

\section{Conclusions}
In summary, we study quantum ring in
noncommutative phase space based on the Seiberg-Witten map and give the
persistent current in mesoscopic rings induced by the effective AB effect. We introduce two variables $\lambda
$ and $\sigma$ as two signatures to detect the noncommutative phase space based on the relationship between the persistent current and the external magnetic flux, which provides an efficient scheme to detect experimentally the noncommutative phase space. It can be expected to further implement this scheme experimentally to give the answer.

The noncommutative algebra is a central concept in quantum mechanics. This noncommutativity leads to many novel phenomena beyond classical physics, such as wave behavior of particles and its coherence, uncertainty of a pair conjugated variables, quantized energy levels, and nonlocal quantum entanglement. These properties provide a lot of ways to understand the fundamental physics of nature, such as elementary particle, gravity and early universe as well as topological invariance in condensed matter physics. It should be a meaningful direction to explore the noncommutativity of space in mesoscopic scale because it is too difficult to directly implement physical experiments in Planck scale.  Our scheme gives a meaningful way or hint to explore noncommutativity of space in mesoscopic scale.


\begin{acknowledgments}
The authors gratefully acknowledge the financial support of the project
from the Fundamental Research Fund for the Central Universities.
\end{acknowledgments}


\bibliographystyle{plain}
\bibliography{apssamp}

\newpage

\begin{figure}
\centering
\resizebox{1\hsize}{!}{\includegraphics*{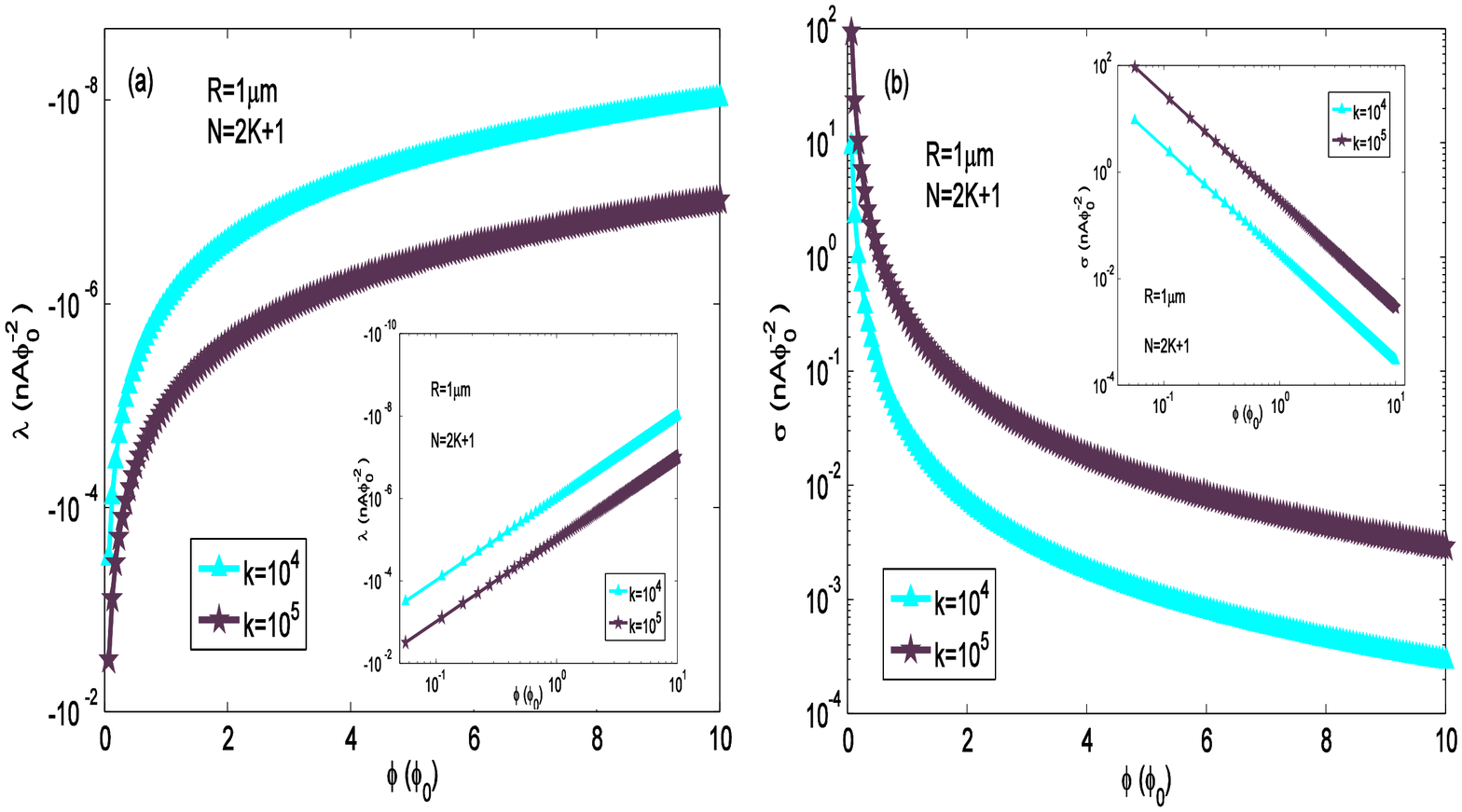}}
\caption{
(Color online): The signatures of noncommutative phase space for odd electron-number rings versus the external magnetic flux. (a) $\log\lambda\sim\phi$ with $\log\lambda\sim\log\phi$ in the inset of (a) as a standard of $-1/\phi$ divergence. (b) $\log\sigma\sim\phi$ with $\log\sigma\sim\log\phi$ in the inset of (b) as a standard of $1/\phi$ divergence. The unit of the magnetic flux is the magnetic flux quantum $\phi_{0}$.}
\label{fig1}
\end{figure}

\begin{figure}
\centering
\resizebox{1\hsize}{!}{\includegraphics*{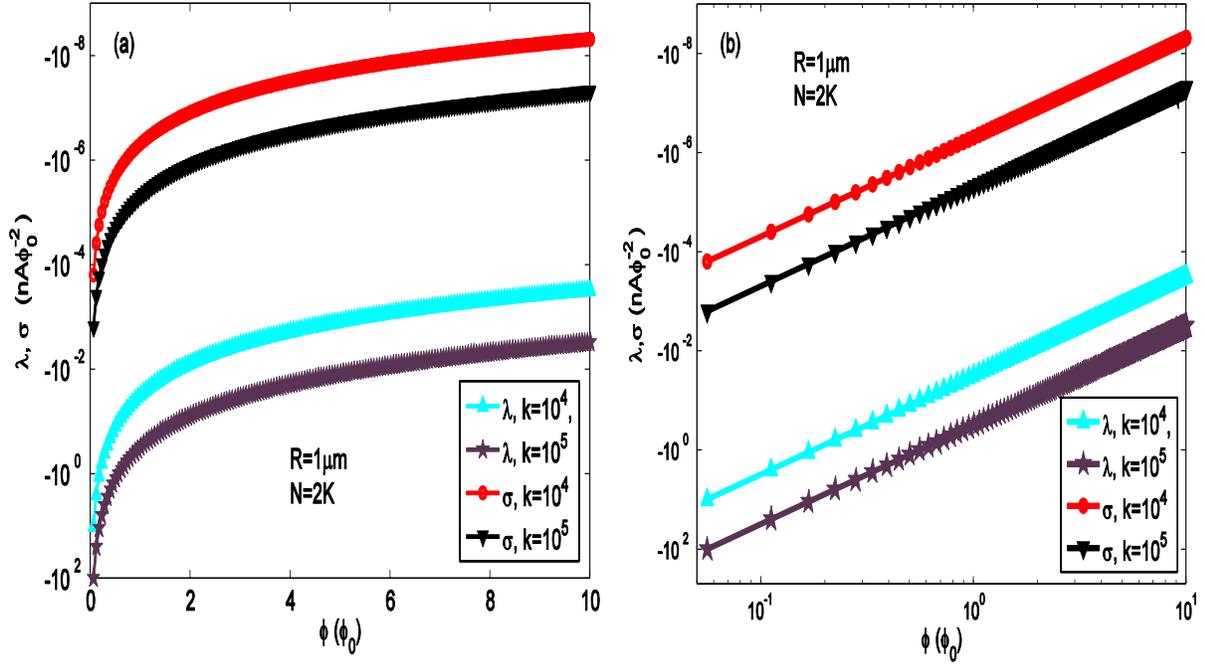}}
\caption{
(Color online): The signatures of noncommutative phase space for even electron-number rings versus the external magnetic flux. (a) $\log\lambda\sim\phi$ and $\log\sigma\sim\phi$ with different electron numbers. (b) $\log\lambda\sim\log\phi$ and $\log\sigma\sim\log\phi$ a standard of $1/\phi$ behavior. The unit of the magnetic flux is the magnetic flux quantum $\phi_{0}$.}
\label{fig2}
\end{figure}

\end{document}